\documentclass[twocolumn,10pt,superscriptaddress]{revtex4-1}
\usepackage{mathtools}
\usepackage{amsfonts}
\usepackage[all]{xy}
\usepackage{tikz}
\usetikzlibrary{arrows, decorations.markings}
\usepackage{subfig}
\usepackage{cmbright}
\usepackage{hyperref}
\usepackage{afterpage}

\newcommand{\beginsupplement}{%
        \setcounter{table}{0}
        \renewcommand{\thetable}{S\arabic{table}}%
        \setcounter{figure}{0}
        \renewcommand{\thefigure}{S\arabic{figure}}%
     }

\begin{document}

\title{Studies of global and local entanglements of individual protein chains using the concept of knotoids}

\author{Dimos Goundaroulis}
\affiliation{Center for Integrative Genomics, University of Lausanne, 1015 Lausanne, Switzerland}
\affiliation{Swiss Institute of Bioinformatics, 1015 Lausanne, Switzerland}
\author{Julien Dorier} 
\affiliation{Center for Integrative Genomics, University of Lausanne, 1015 Lausanne, Switzerland}
\affiliation{Vital-IT, SIB Swiss Institute of Bioinformatics, 1015 Lausanne, Switzerland}
\author{Fabrizio Benedetti}
\affiliation{Center for Integrative Genomics, University of Lausanne, 1015 Lausanne, Switzerland}
\affiliation{Vital-IT, SIB Swiss Institute of Bioinformatics, 1015 Lausanne, Switzerland}
\author{Andrzej Stasiak}
\affiliation{Center for Integrative Genomics, University of Lausanne, 1015 Lausanne, Switzerland}
\affiliation{Swiss Institute of Bioinformatics, 1015 Lausanne, Switzerland}
\affiliation{To whom correspondence should be addressed. E-mail: andrzej.stasiak@unil.ch}





\begin{abstract}
We study here global and local entanglements of open protein chains by implementing the concept of knotoids. Knotoids have been introduced in 2012 by Vladimir Turaev as a generalization of knots in 3-dimensional space. More precisely, knotoids are diagrams representing projections of open curves in 3D space, in contrast to knot diagrams which represent projections of closed curves in 3D space. The intrinsic difference with classical knot theory is that the generalization provided by knotoids admits non-trivial topological entanglement of the open curves provided that their geometry is frozen as it is the case for crystallized proteins. Consequently, our approach doesn't require the closure of chains into loops which implies that the geometry of analysed chains does not need to be changed by closure in order to characterize their topology. Our study revealed that the knotoid approach detects protein regions that were classified earlier as knotted and also new, topologically interesting regions that we classify as pre-knotted.
\end{abstract}

\maketitle

\section*{Introduction}

Since early observations of topological entanglements in biopolymers  \cite{Dean, Con, Mans}, biology became one of practical fields of application of knot theory. For instance, the appearance of particular knot types in unknotted circular DNA molecules that served as a substrate of various enzymes helped to determine the molecular mechanism of these enzymes \cite{spengler, sumners, crisona, buck, olorun}. Likewise, the analysis of DNA knots formed inside bacteriophage capsids permitted to elucidate how the densely packed DNA is arranged within heads of bacteriophages \cite{arsuaga2005dna, marenduzzo2009dna, reith2012effective}. 
From a mathematical point of view, a knot is a closed curve in 3-dimensional Euclidean space that does not intersect itself anywhere and can be continuously deformed as if it was made out of rubber \cite{Adams}.  For closed curves, any continuous deformation maintains the original topology. Thus, for example, a trefoil knot will always be a trefoil knot upon continuous deformation.
In case of proteins, which are open linear chains with complex geometry, a continuous deformation can convert them into a straight open chain. Therefore from an orthodox topological point of view all proteins with open chains are unknotted. However, to analyse the topology of protein chains it makes sense to treat their configurations as rigid and thus not able to undergo any continuous deformation. In fact, proteins can undergo some internal motion but crystallized forms are essentially rigid. Here we work mostly with coordinates of such proteins. By introducing the condition that proteins are rigid, the question of protein knottedness becomes interesting and treatable. Early approaches required though closure of the protein polypeptide chain before the analysis of protein topology \cite{Mans}. Of course, how to close the protein is a nontrivial question and several methods were proposed \cite{millett2013}. The problem is that different methods of closure may lead to different knot types being associated with a given protein. Closure was required in the past since available mathematical tools that can be used to determine the knot type could only analyse closed curves. Interesting advances opening the possibility to analyse topology of open curves came with the discovery of concept of knotoids and with the mathematical tools permitting their analysis \cite{tu, guka}. Using the knotoid concept, the topology of open chains can be analyzed using just projections of these chains without the formal need to close them. We use here the knotoid approach to analyse knottedness of entire protein chains, as well as of their all possible subchains.

\section*{A primer in knot theory}

\begin{figure}[tbhp]
\centering
\includegraphics[width=.25\linewidth]{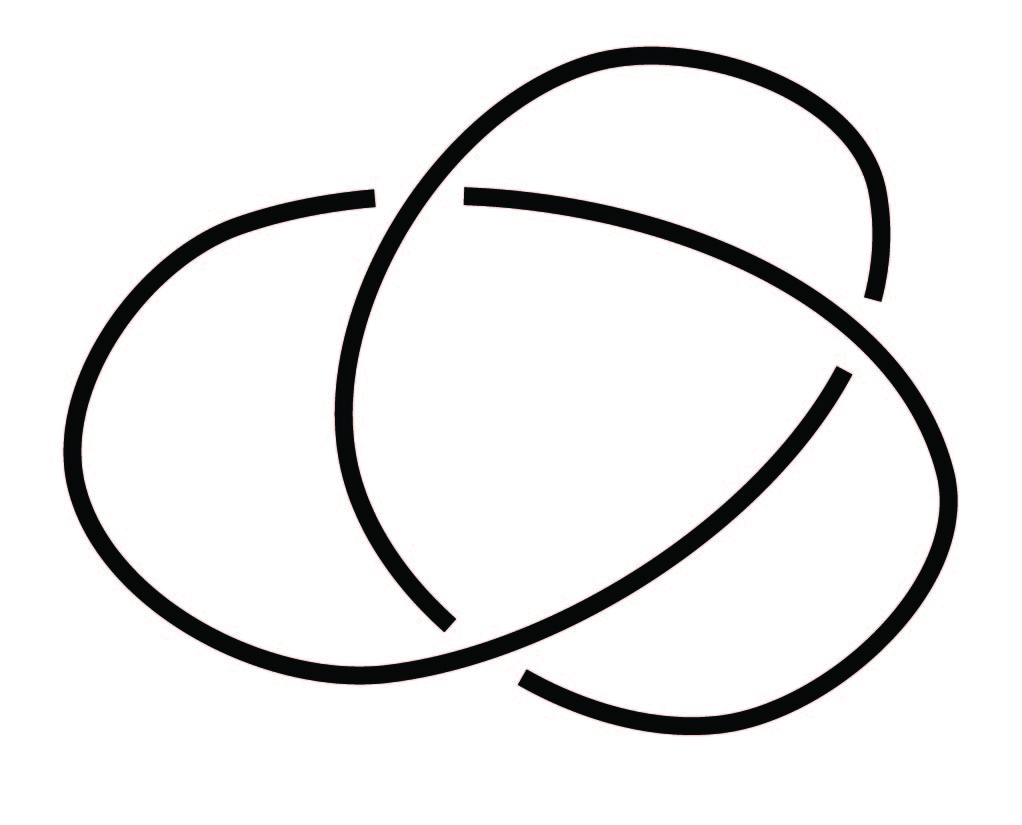}
\caption{A diagram of the trefoil knot.}
\label{fig:trefoil}
\end{figure}

\begin{figure}[tbhp]
\centering
\includegraphics[width=1\linewidth]{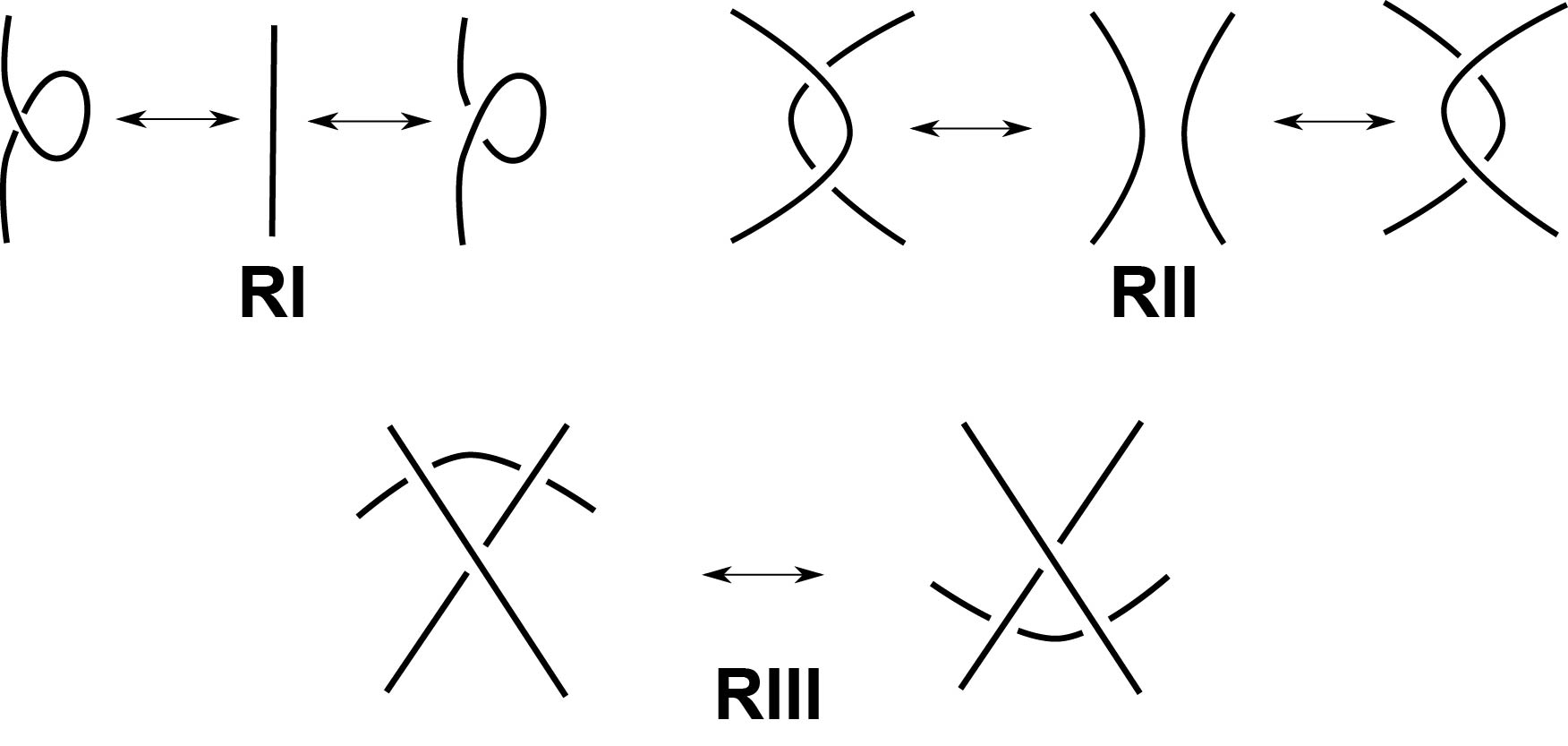}
\caption{Reidemeister moves. Top Row: Reidemeister Moves I and II. Bottom Row: Reidemeister Move III}
\label{fig:rmoves}
\end{figure}

Mathematical knots are usually studied through their diagrams. A knot diagram is obtained by projecting the knot onto a plane in such a way so that only double points are allowed in the projection. A double point or a {\it crossing} of the knot diagram, is  a point of the diagram where two arcs of the knot cross transversely one another (see Fig.~\ref{fig:trefoil}). The crossings also carry the extra information that indicates the undercrossing arc. In this way, one can always reconstruct of the knot in the 3D space \cite{kauphys}.
Two knots are equivalent, if they can be continuously deformed to one another without allowing cutting and regluing of the knot. On the level of diagrams, this equivalence is proven if the two diagrams can be transformed to one another by a finite sequence of three elementary diagram moves, known as the Reidemeister moves  (see Figure~\ref{fig:rmoves}) and planar isotopy i.e. stretching, shrinking, bending or straightening of portions of the diagram in the plane so that the underlying structure is preserved (see for example \cite{Adams, kauphys, kau1}). However, finding out whether there is a combination of Reidemeister moves and planar isotopy transformation that can convert two diagrams into each other may be a very difficult task, especially if diagrams have many crossings. In such cases, more sophisticated tools are required. Such tools are the knot invariants, which are functions defined on the set of all knots that assign the same value to equivalent knots. The most well-known knot invariants are the knot polynomials such as the Alexander polynomial  \cite{alex}, the Jones polynomial \cite{jo} and its generalization the
Homflypt polynomial \cite{homfly, pt}.

\section*{Knots in proteins}
Numerous studies conducted in the last twenty-five years have revealed the existence of a large number of
proteins whose main chain fold into non-trivial topologies, a fact that implies the presence of knots in their
conformation \cite{Mans, taylor, virnau, mallam2006probing, mallam2008exploring, yeates2007knotted, sulkowska2012energy}. The precise nature of the structural and functional advantages created by the
presence of knots in the protein backbone is a subject of high interest from both experimental and theoretical
point of view. It has been conjectured that these non-trivial topologies provide a stabilizing function that can
act by holding together certain protein domains \cite{yeates2007knotted, sulkowska2008stabilizing, virnau, Dabr} . However, the precise structural and functional advantages
provided by the presence of knots is uncertain in the majority of the cases.
To better understand this open problem, several efforts have been made towards the
characterization and classification of the protein chains based on their knot type \cite{sulk}. 
This characterization required though an important departure or even apostasy from the orthodox knot theory that consists of accepting that linear chains can be knotted. This contradicted
the central axiom of knot theory where any open arc, no matter the degree of entanglement, is topologically
equivalent to a straight line since any open arc can be continuously deformed to a straight line   \cite{Adams}. 
However, when the continuous deformation is not allowed, the determination of knottedness of a given open curve with frozen geometry starts to make sense and is mathematically challenging. In fact, proteins in their native folded structure are frequently quite rigid and show only limited internal motion. Therefore analysis of their knottedness is done for their open chains with fixed geometry.

Having established the above assumptions, a procedure to capture the knotting type has to be chosen. Until recently, characterization of knottedness of proteins required closure of protein chains since available knot invariants could only be calculated for closed curves. In this context, various methodologies of chain closure have been proposed, both deterministic  (for example \cite{taylor,
virnau}) and probabilistic (e.g. \cite{Mans, lua, millett, sulk, jamroz}). All deterministic methods deal with the choice of a closure for the open chain, and setting the condition of choice is somewhat arbitrary. Depending on the choice of closure, different knot types can be associated to the same protein. To avoid this arbitrarity and a possible bias, probabilistic methods were introduced that use unbiased multiple closures to detect the most likely knot type of a linear chain with a given geometry \cite{Mans, lua, millett, sulk, jamroz}. An example of such a method is the following: the chain is placed near the
center of a large sphere. Then, both ends of the chain are extended towards the direction of a randomly
chosen point on the sphere where they are joined. This procedure, after multiple repetitions, produces a
spectrum of knots that are associated to the given linear chain \cite{millett}.
After every closure, either by deterministic or probabilistic methods, formed knots' identification is achieved via computation of a knot polynomial. Note that knot polynomials are not complete invariants, in the sense that there are pairs of knots that cannot be distinguished (e.g. mutant knots). However, all knots that are encountered within the backbone of a protein are relatively simple and thus they can be identified by polynomial invariants of knots such as the Jones polynomial. 

All approaches that require protein chain closure, deterministic and probabilistic, suffer though from a formal problem. The configurations that are analyzed for knotting are not the original configurations of studied proteins but configurations that were changed i.e. deformed by the addition of the closing parts.

We describe here a method that allows us to characterize knottedness of unperturbed configurations of proteins using the concept of knotoids. That concept is explicitly used to analyze diagrams resulting from the projections of open chains. 

\section*{Knotoids as an extension of knot theory}

Knotoids were first introduced by V. Turaev in \cite{tu} and have been studied further by L. Kauffman and N.
G\"ug\"umc\"u in \cite{guka}. In brief, they are equivalence classes of open ended knot diagrams that generalize the notion of a 1-1 tangle (or a long knot) since they allow the endpoints to be in different regions of the diagram, and thus they provide a rigorous definition for open knots. In this way, a new diagrammatic theory is formed which is an extension
of the classical knot theory \cite{guka}. 

\begin{figure}[ht]
\centering
\includegraphics[width=1\linewidth]{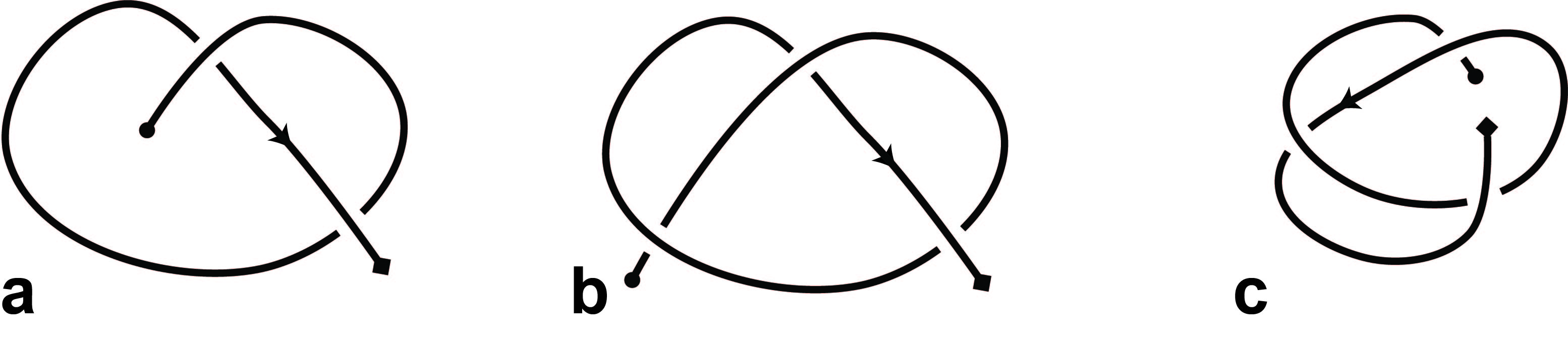}
\caption{Types of knotoids. (a) A non-trivial pure knotoid diagram with two crossings, (b) A knot-type knotoid with three crossings and (c) A knot-type knotoid with three crossings.}
\label{fig:knotoid}
\end{figure}

Knotoid diagrams were originally defined in the 2-sphere $S^2$ however, their definition can be extended to $\mathbb{R}^2$. A knotoid diagram is defined as the generic immersion of the closed unit interval in $S^2$ whose only singularities are transversal double
points endowed with over/undercrossing data. The images of 0 and 1 under this immersion are
called {\it the tail} and {\it the head} of the knotoid diagram respectively. These two points are distinct from
each other and from the double points; they are also called {\it the end points} of the knotoid diagram. Every
knotoid diagram comes with an orientation that goes from the tail to the head. The double points of the
knotoid diagram are called {\it crossings} \cite{tu} (see Figure~\ref{fig:knotoid}).

\begin{figure}[ht]
\centering
\includegraphics[width=1\linewidth]{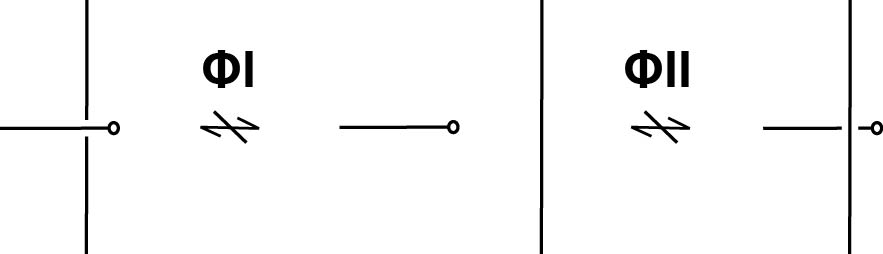}
\caption{The forbidden moves. Crossing under ($\Phi$I) and over ($\Phi$II) an arc adjacent to an endpoint is prohibited.}
\label{fig:forbmoves}
\end{figure}

Two knotoid diagrams are considered equivalent if and only if they differ by a finite sequence of the Reidemeister moves that modify the diagrams within a small disk but which do not utilize the endpoints. The corresponding equivalence classes are called {\it knotoids}.  Since it is forbidden to move diagram portions over or under the end points, a non-trivial knotoid diagram cannot reduce to the trivial one (see Figure~\ref{fig:forbmoves}).  Any knotoid that has both endpoints in the same local region of $S^2$ is called {\it a knot-type} knotoid and they are denoted by $k^\circ$, where $k$ is the corresponding entry in the knotoid table. All other knotoids are called {\it pure} or {\it proper knotoids}. Many knot invariants, like the Kauffman bracket and the Jones polynomial extend to the case of knotoids in a natural way \cite{tu, guka}.

\section*{Knotoids, open curves and protein chains}

\begin{figure*}[ht]
\centering
 \includegraphics[width=1\linewidth]{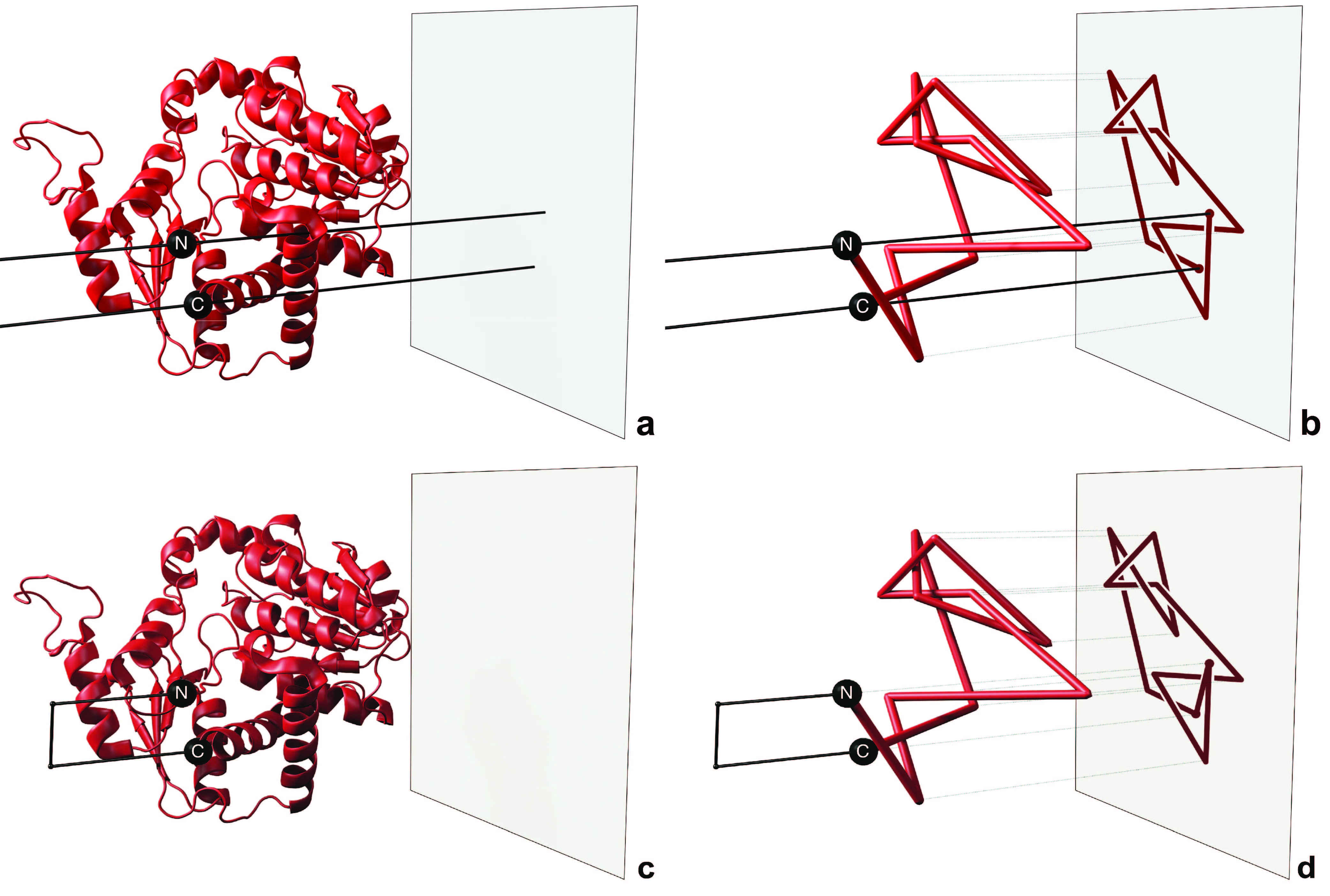}
  \caption{Projection of a protein chain using two different techniques. Top Row: Knotoids technique. (a) The two black infinite lines pass through the N and C termini of the protein chain and are perpendicular to a chosen plane. (b) The two infinite lines pass through the N and C termini of the reduced protein backbone are perpendicular to a chosen plane. Bottom Row: Stochastic closure technique. (c) A choice of closing direction and the two rays extending from the termini towards that direction. The ends of the two rays are connected when they exit the sphere that contains the protein chain. (d) The resulting knot diagram.}
  \label{fig:proteinlines}
\centering
\end{figure*}

In this section we shall discuss the connection between knotoid diagrams and oriented curves in the 3-dimensional space. A knotoid diagram represents an open oriented curve in $\mathbb{R}^3$ if it is in the equivalence class of the knotoid that corresponds to the generic projection of the curve to some plane. Consider now a smooth curve which lies inside a large ball in $\mathbb{R}^3$.  Each point of that ball points towards a generic projection to a plane that lies outside the ball in $\mathbb{R}^3$. The two end points of the curve determine two parallel lines, each one passing through one endpoint. Both lines are perpendicular to the plane that corresponds to the generic projection. By considering the generic projection of the curve to the plane along the lines together with the information of the overpassing and underpassing arcs, one obtains a knotoid diagram in $\mathbb{R}^2$ \cite{guka} (see Fig.~\ref{fig:proteinlines}). The resulting knotoid diagrams and their knotoid types depend on the choice of a projection plane. However, to characterize knottedness of open curves such as represented by structures of proteins, we characterize the spectrum of knotoids observed when a given open curve is projected in all possible directions equisampling the sphere. The most frequently observed knotoid type is then associated with the given structure as its dominant knotoid type.

Although the knotoid approach allows us to study knottedness of protein chains without any deformations of the chains, the knotoid types formed by projections with many crossings are difficult to determine as the computation of Jones polynomials becomes too time demanding. In such cases we simplify knotoids diagrams by actually doing triangle elimination moves on 3D configurations of analysed proteins \cite{koniaris}. Importantly, the knotoid type is not changed when one does not permit  triangle elimination moves to pass through the two infinite lines that are perpendicular to the projection plane and which go through the ends of the linear chains (see Fig.~\ref{fig:proteinlines}A, B). 

One can now define a measure for the entanglement of a smooth open curve in $\mathbb{R}^3$ by  analyzing many directions of projections equisampling the sphere. More precisely, the set of all knotoid equivalence classes that are obtained from generic projections of the curve in question to different planes in $\mathbb{R}^3$ is the {\it measure of knottedness} of the curve.  The knotoid with the highest number of occurrences in this measure shall be called {\it dominant} and in this study we are mainly interested in the determination of the dominant knotoid type for a given protein chain.

If the dominant knotoid that appears in the measure is the trivial one (or the {\it unknotoid}) then the chain is considered as topologically not entangled. Otherwise, the chains are considered as topologically entangled and we characterize them by determining the dominant knotoid type resulting from projections of a given chain.

\section*{Results}
\subsection*{Global entanglement: Knotoids versus stochastic closure}

\begin{figure*}[ht]
\centering
\includegraphics[width=1\linewidth]{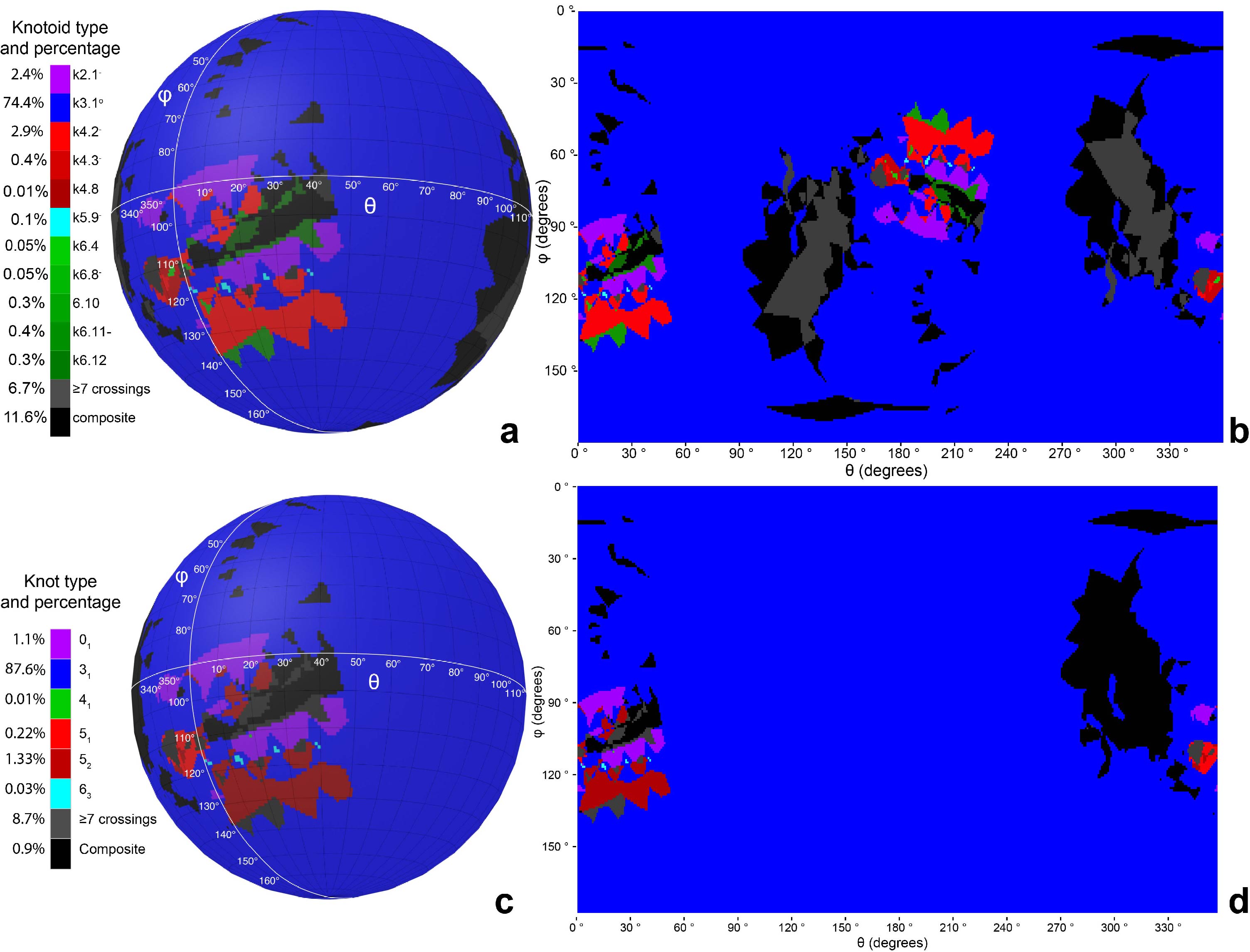}
  \caption{Projection Maps. Top Row: The knotoid technique for the protein 3KZN. Bottom Row: The stochastic closure technique for the protein 3KZN. Maps were created with the open-source softwares Blender 2.78 and and R 3.3.2 (respectively \url{https://www.blender.org/} and \url{https://www.r-project.org/}).}
  \label{fig:uniclo}
\end{figure*}

Our first aim is to study the global entanglement of protein chains. The protein structure we analyze here is this of bacterial N-acetylornithine transcarbamoylase with the PDB entry 3KZN and it is known to form an open trefoil knot. Figure~\ref{fig:uniclo}A shows what types of knotoids are obtained when the protein is projected along a large number (80,000) of  directions. The results are presented as a map that identifies territories on the surface of the sphere, where each distinct territory (with a given color) encloses directions of projections resulting in a given knotoid. Depending on the original orientation of the protein, the map produced on the surface of the sphere can rotate but the fraction of projections that result in a given knotoid stays always the same for a given protein structure.  The 3KZN protein forms a deep and relatively tight, open trefoil knot \cite{Dabr}. As could be expected for that particular protein, the great majority of projections resulted in diagrams, which upon topology conserving moves can be simplified to a knot-type knotoid diagram with 3 crossings. The adopted here topological notation of that knotoid is ${\rm k}3.1^\circ$. The superscript $^\circ$ indicates that it is a knot-type knotoid, which means that its diagram can be closed without crossing other parts of the diagram. In this case such a closure will result in $3_1$ knot. In fact, more than 75\% of projections of 3KZN protein resulted in ${\rm k}3.1^\circ$ knotoids. The territories enclosing directions of projections resulting in ${\rm k}3.1^\circ$ knotoid are indicated with the blue color on Figure~\ref{fig:uniclo}A. Figure~\ref{fig:uniclo}B shows the equirectangular projection of the spherical map shown in Figure~\ref{fig:uniclo}A. Although the equirectangular projections distort the relative sizes of distinct knotoid territories we do not use them here to evaluate the relative area of these territories but to show that antipodal directions of projections give the same knotoid types. Looking at the knotoid maps of 3KZN protein (Fig.\ref{fig:uniclo}A, B), we can see that in addition to projections resulting in knotoid ${\rm k}3.1^\circ$ there are projections resulting in formation of many other types of knotoids, including some with more than 6 crossings in their minimal crossing diagrams (Figures S1 and S2 presents the types of knotoids resulting from the projection of studied proteins). Interestingly, there were no trivial knotoids ${\rm k}0.1$. However, we did observe the simplest nontrivial knotoids with the notation ${\rm k}2.1$. These knotoids have just two crossings in their diagrams (see Fig.~\ref{fig:knotoid}A) but can't be continuously deformed into trivial knotoid ${\rm k}0.1$. Using knotoid approach to characterize topology of proteins, we associate the most frequent knotoid type to a given protein or its subchain.  Table~\ref{domtabl} lists the dominant knotoid types and the fraction of random direction that give rise to the corresponding knotoid type for 14 proteins that are known to form open knots \cite{sulk}.

\begin{table}[ht]
\centering
\begin{tabular}{llc}
\hline
protein & knotoid & percentage \\ \hline
1XD3    &    ${\rm k}5.2^{\circ -}$    & 41.6\%     \\
1YRL    &    ${\rm k}4.1^\circ$     & 49.9\%     \\
1YVE    &     ${\rm k}4.1^\circ$    & 42.4\%     \\
2AXC    &     ${\rm k}0.1$    & 89.1\%     \\
2JLO    &     ${\rm k}0.1$    & 81.7\%     \\
2OOL    &    ${\rm k}4.1^\circ$     & 65.9\%     \\
3BJX    &     ${\rm k}6.1^\circ$    & 42.2\%     \\
3C2W    &     ${\rm k}4.1^\circ$    & 68.1\%     \\
3DH4    &   ${\rm k}0.1$      & 68.2\%     \\
3FR8    &    ${\rm k}4.1^\circ$     & 38.1\%     \\
3IRT    &     ${\rm k}5.2^{\circ -}$    & 40.0\%       \\
3KZN & ${\rm k}3.1^\circ$ & 75.6\%	\\
3L1L    &    ${\rm k}0.1$     & 80.6\%     \\
3NCY    &   ${\rm k}0.1$      & 77\%      
\end{tabular}
\caption{Proteins and their dominant knotoids}
\label{domtabl}
\end{table}

It is interesting to compare the characterization of protein chain topology using knotoid approach, presented above (Figure~\ref{fig:uniclo}A, B), with the stochastic closure technique, described earlier (see Fig.~\ref{fig:uniclo}C, D)  \cite{sulk}. Stochastic closure technique is formally equivalent to the first phase of knotoid approach, i.e. the protein chain is placed in the centre of a big sphere and is projected along random directions on the surface of the enclosing sphere. From here on, the closure approach diverge from the knotoid approach. In the stochastic closure approach the projection-derived diagrams, containing the information which segments were above and which under, are closed with a straight segment, where the closing segment passes over all other segments it crosses with on the diagram (see Fig.~\ref{fig:proteinlines}C, D). The knot type of the diagram gets fixed upon closure and its type can be directly determined by the calculation of a knot invariant such as the Jones polynomial. To facilitate the computation of knot invariants the diagrams may be simplified by triangle elimination moves and by Reidemeister moves. For purpose of better comparison between knotoid and knotting approach we use the same directions of projections for both approaches. Looking at figure~\ref{fig:uniclo}A, B and~\ref{fig:uniclo}C, D it is not difficult to observe that the knotoid approach involves a wider spectrum of knotoid types when compared the spectrum of knots, even if we group all knotoids or knots with more than 6 crossings into one category. Additionally, we observe that in both cases the dominant knot type involves structures with three crossings; the trefoil knot for the case of the stochastic closure and the knot-type knotoid ${\rm k}3.1^\circ$. Further, one can see on the knotoid map that all territories corresponding to knot type knotoids (in this case ${\rm k}3.1^\circ$) are contained within corresponding knot type territories on the knot map. The situation is different though for proper types of knotoids as only one of their antipodal territories is carried through to the knot map with the appropriate change of topological notation.

One may recover the knotoid map from the knotting map by ``twinning'' of all territories of non-dominant knots, i.e. by adding the antipodal territories to territories of non-dominant knots. While the overall shapes of territories which indicate projection directions that generate diagrams of non-dominant knots and knotoids are maintained. The knotoid territories are divided into a larger number of different topological subclasses. This indicates that knotoid approach provides a finer topological distinction between various topological states than knotting approach.
Comparing the knotoid approach to the knotting approach, we can see that one obtains similar but not identical information about the topology of analysed chains by using these two methods. It is comforting to see that the method that does not invoke closure of analysed configurations captures similar topological features as a method that always closes analysed trajectories.

 Very recently Alexander et al., used virtual knot formalism to analyse configurations of knotted proteins. Although virtual knot formalism requires closure of the diagrams derived from projections of open knots, the virtual closure imposes very similar limitations on topology of the virtual knot diagrams as the end points of knotoid diagrams. Therefore the methods of studying the topology of a protein chain using the knotoids approach and the virtual knots approach \cite{altade} produce very similar results.

\subsection*{Local entanglement: Slipknotoids and knotted cores}

\begin{figure*}[ht]
  \centering
  \includegraphics[width=1\textwidth]{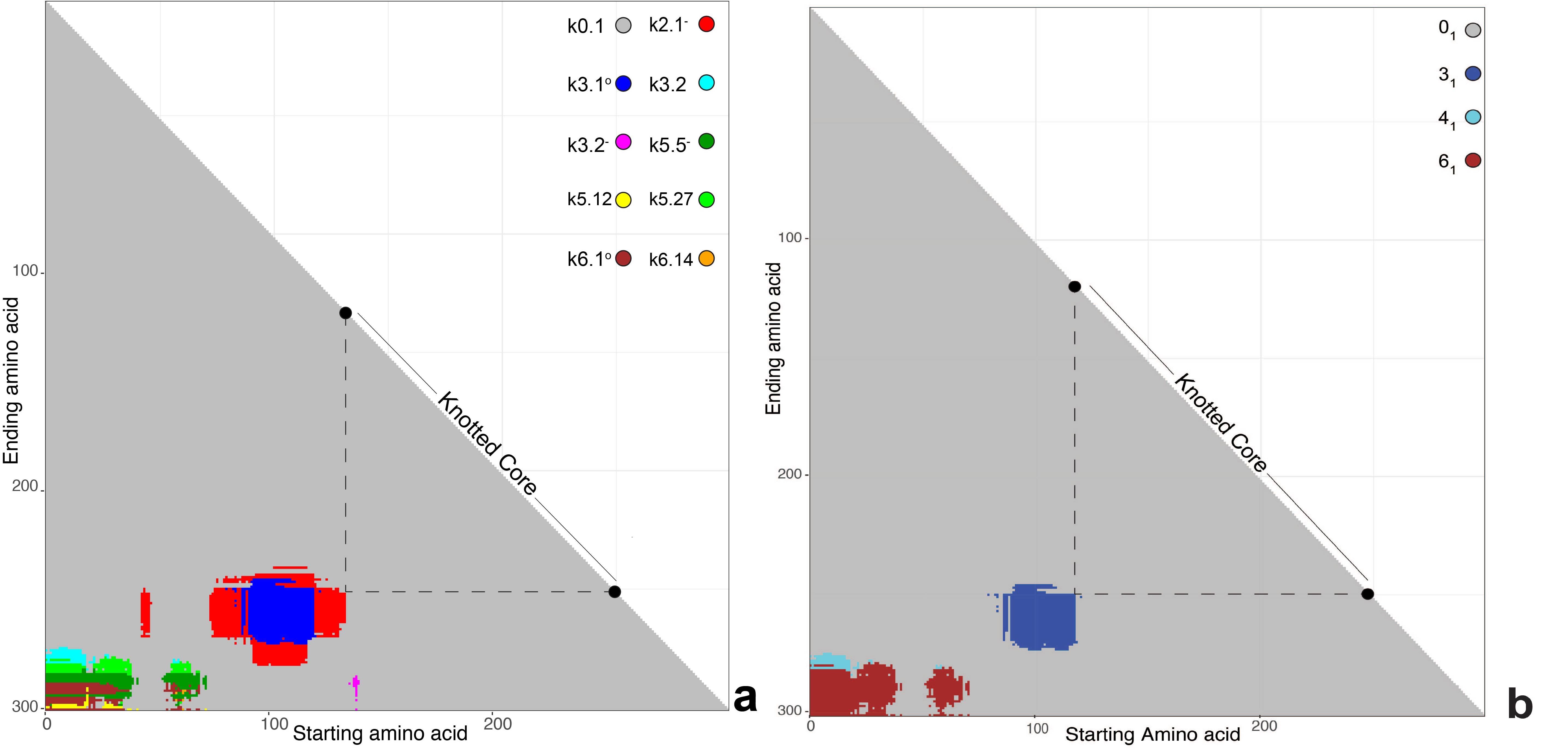}
  \caption{Fingerprints. (a) The knotoid fingerprint of DehI. (b) The knot fingerprint of DehI.}
  \label{fig:3BJXmats}
\end{figure*}

Characterization of the global knottedness of proteins by knotoid or chain closure approach tells us only what type of open knot forms a given entire protein. This method does not inform us where the knotted portion is located within the protein structure, how large is the knotted core or whether a given polypeptide chain contains subknots, which are less complex knots located within more complex knots \cite{rawdon2015subknots}. This information can be provided though when one analyses knottedness of every possible subchain of a given protein \cite{yeates, taylor95}. The analysis of every subchain provides what is known as the {\it knotting fingerprint} of a given protein, which is usually presented in a form of triangular matrix where every entry in the matrix informs what is the dominant knot type for a given subchain \cite{sulk}. The dominant knotoid types are indicated in the provided coloured scheme (see Fig.~\ref{fig:3BJXmats}A). We decided to compare knotting fingerprints of several knotted proteins with their knotoid fingertprints. Since for a given number of crossings in a minimal crossing diagram there are more types of knotoids than of knots, knotoid approach can provide a finer characterization of protein topology than the knotting approach necessitating chain closure.

Given a protein chain, we start studying its local entanglement behaviour by clipping the chain one alpha-Carbon atom at a time, starting from either of the termini of the chain. Each time we obtain a shorter chain, which we analyze in terms of the knotoid technique that is, by projecting the trimmed chain along 1000 random directions and then computing the Jones polynomial of each projection. Notice that as we trim the chain, we observe that the knotoids types change. The shortest subchains that form a given knotoid type are cores of that particular knotoid and their size and position is indicated by entries of a given colour that are closest to the diagonal of the matrix.

Our working example here is the protein DehI ($\alpha$-haloacid dehalogenase) whose backbone forms a $6_1$ knot, the most complex protein knot known so far \cite{bol}. In \cite{sulk} all subchains of DehI were analyzed using the stochastic closure technique and it was observed that, in addition to the $6_1$ dominant knot formed by the entire protein, smaller subchains formed $4_1$ and $3_1$ knots. Interestingly, after analyzing DehI using the knotoid approach and comparing the resulting knotoid fingerprint to its knot fingerprint (see Fig.~\ref{fig:3BJXmats}A, B) we observe that   the knotoid approach exhibits a much richer diversity in terms of the different topological forms.

Going into further detail, the dominant knotoid type of the entire chain is the knot-type knotoid 6.1. Recall that knot-type knotoids have both endpoints in the same region of the plane and so they always yield the same knot regardless of how one chooses to close the knotoid diagram but without introducing additional crossings. Returning to the comparison, of knotoid and knotting figerprints, we can see that the regions of the knotoid fingerprint that correspond to the trivial knotoid carry through to the matrix of the knot fingerprint as trivial knots. Furthermore, the regions of non-trivial knots of the knot fingerprint of DehI are contained within regions of the knotoid fingerprint. In addition, there are new regions  in  knotoid fingerprints that correspond to non-trivial proper knotoids that either border regions of knot types knotoids (e.g. the thick regions of of $2.1^-$ knotoids that encircle the $3.1^\circ$) or show up as islands within trivial knotoids (the small slices of  $2.1^-$ and $3.2^-$).  These regions on knotoid fingerprints that are not visible on knotting fingerprints correspond to polypeptide portions that are not completely knotted and we propose to call them {\it pre-knots}.  

\begin{figure*}[ht]
  \centering
\includegraphics[width=.7\linewidth]{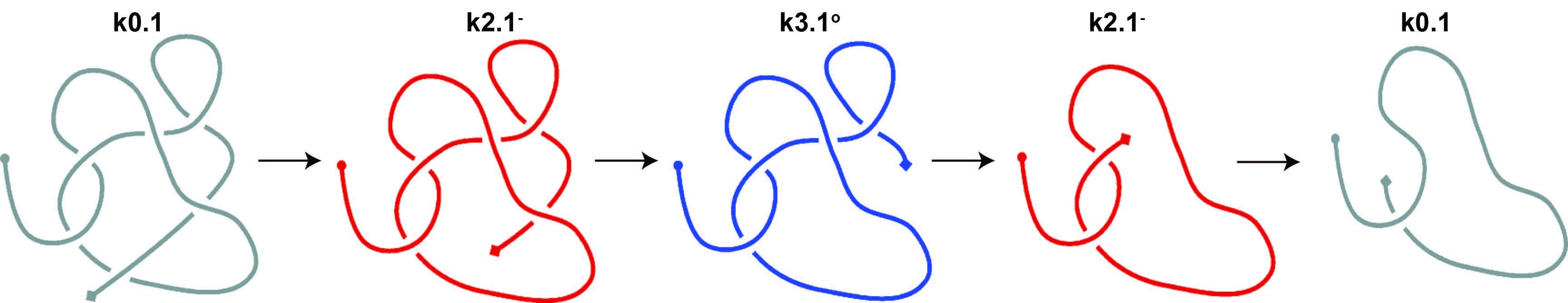}
\caption{Instance of the trimming process of protein 3BJX. Notice that only one end of projected chain is progressively trimmed. The knotoid type change is indicated with the change of the color of the diagram and the used colors correspond to these used to indicate knotoid types in Figure~\ref{fig:3BJXmats}.}
\label{fig:tref_crop}
\end{figure*}

Figure~\ref{fig:tref_crop} explains how by progressively trimming one end of an entangled subchain that forms a trefoil slipknot one can pass from trivial knotoid to pure knotoid 2.1, then to knot type knotoid $3.1^\circ$, then to pure knotoid 2.1 and finally to trivial knotoid again. The starting diagram in Figure~\ref{fig:tref_crop} is based on \cite{sulk}.

It is interesting to compare the size of the knotted core to the size of knotoid core. The size of   knotted core in knotting fingerprint corresponds to the size of the knot type knotoid on knotoid fingerprint. Pure knotoids though have smaller cores than knot type knotoids. This is a trivial consequence of the fact that trimming the subchain that forms a knot type knotoid one usually passes first into a subchain that forms pure knotoid before passing to a subchain that forms trivial knotoid.

\section*{Discussion}

In order to study the topology of a protein chain there are two options so far. The first one is to study the protein as a closed chain, so one has to determine a way to close the chain and then study the resulting knot. The second option would be to study all possible projections of the open chain resulting in diagrams of knotoids. 

The advantage of using the knotoids approach is the increased sensitivity of entanglement detection. In this paper we have shown that the immediate impact of this can be observed in the subchain analysis and local entanglement study of a protein chain. To be more precise, our study indicated that the knotoids approach produces more refined fingerprints of the protein chains. Additionally, the knotoids method detects in higher detail the minimal length of the chain that can remain tangled giving thus  a more accurate overview of the entanglement pattern of each protein chain. Knotoid fingerprints detect also the protein regions with topological entanglement that is not detected by knotting fingerprints. We propose to call these protein portions as pre-knots. Figures S3-S5 show knotoid fingerprints of 12 other proteins that were analyzed with knotting fingerprint by Su{\l}kowska et al.

\section*{Methods}
For our analysis, we import for each protein its structure from the PDB and convert it to $xyz$-coordinates. We reconstruct the protein chain using only $C^\alpha$ atoms and then choose its direction of projection. Once the direction is chosen we perform triangle simplification moves \cite{koniaris} that do not pass through two lines that are parallel to the direction of projection and which go through the end points of the reconstructed protein chain. This procedure keeps the knotoid type while reducing the number of nugatory crossing, which in turn facilitates the computation of Jones polynomials. We project the protein chain on a plane perpendicular to the projection direction and evaluate the knotoid diagram. Repeated applications of the three Reidemeister moves allows us to reduce the number of crossings of the corresponding knotoid diagram.

For each projection we compute its Jones polynomial for the case of knotoids \cite{tu, guka} in the following way.  We smooth each crossing of the knotoid diagram $K$ using the following rules:
\begin{eqnarray}
&&\langle \raisebox{-.1cm}{\begin{tikzpicture}[scale=.2]
\draw [line width=0.35mm]  (-1,-1)-- (-0.22,-0.22);
\draw  [line width=0.35mm ](-1,1)--(0,0);
\draw  [line width=0.35mm] (0.22,0.22) -- (1,1);
\draw [line width=0.35mm]   (0,0) -- +(1,-1);
\end{tikzpicture}}\, \rangle =A \langle \, \raisebox{-.07cm}{\begin{tikzpicture}[scale=.2, mydeco/.style = {decoration = {markings, mark = at position #1 with {\arrow{>}}}}]
\draw [line width=0.35mm] plot [smooth, tension=2] coordinates { (-1,.8) (0, 0.5) (1,.8)};
\draw [ line width=0.35mm] plot [smooth, tension=2] coordinates { (-1,-.8) (0, -0.5) (1,-.8)};
\end{tikzpicture}}\, \rangle   + A^{-1} \,\langle\, \raisebox{-.1cm}{\begin{tikzpicture}[scale=.2]
 \draw [ line width=0.35mm] plot [smooth, tension=2] coordinates { (-1,-1) (-0.3, 0) (-1,1)};
 \draw [ line width=0.35mm] plot [smooth, tension=2] coordinates { (1,-1) (0.3, 0) (1,1)};
 \end{tikzpicture}}\, \rangle \label{regbracket1} \\
&&\langle K \sqcup \bigcirc \rangle = \left (-A^2 - A^{-2}\right ) \langle K \rangle \label{regbracket2}\\
&&  \langle \, \raisebox{.07cm}{\begin{tikzpicture}[scale=.1, baseline]
\draw[line width=0.35mm] 
  (1,0) 
    .. controls (3,2) and (5,-2) .. 
  (7,0);  
\draw[black,fill=black] (1,0) circle (2ex);
\draw[black,fill=black] (7,0) circle (2ex);
\end{tikzpicture}}\, \rangle  = 1 \label{regbracket3}
\end{eqnarray}
Equations~\ref{regbracket1}-\ref{regbracket3} comprise the extension of the Kauffman bracket polynomial to the case of knotoids \cite{guka}. The diagrams involved in \eqref{regbracket1} are identical except at one crossing. The second rule means that whenever we have a disjoint circle in a state, we can remove it and multiply by $(A^2 - A^{-2})$.  The Jones polynomial for a knotoid diagram $K$ is then given by the following:
\begin{equation}\label{jones}
f_K(A) = (-A^3)^{- {\rm wr}(K)} \langle K \rangle,
\end{equation}
where $\langle K \rangle$ is the Kauffman bracket polynomial of $K$ and ${\rm wr}(K)$ is the writhe of $K$. The Jones polynomial for knotoids is computed by a routine that follows \cite{drorknot}. For the subchain analysis, we remove one bead at a time starting from the $C$-terminus of the chain and we apply to the resulting chain the above procedure. Once we identify the dominant knotoid for each subchain, we proceed and create the knotoid fingerprint using \texttt{R}.

\section*{Acknowledgements}

This work was funded in part by the Leverhulme Trust (RP2013-K-017 to A.S.) and by the Swiss National Science Foundation (31003A-138267 to A.S.)

\section*{Author contributions statement}

A.S. and D.G. designed research; D.G., J.D., F.B. performed the research; D.G., J.D., F.B. and A.S. analyzed the data; D.G. and A.S. wrote the paper. All authors reviewed the manuscript.



\onecolumngrid
\beginsupplement

\subsection*{Supplementary Information}

\begin{figure*}[!thbp]
  \centering
\includegraphics[width=.9\textwidth]{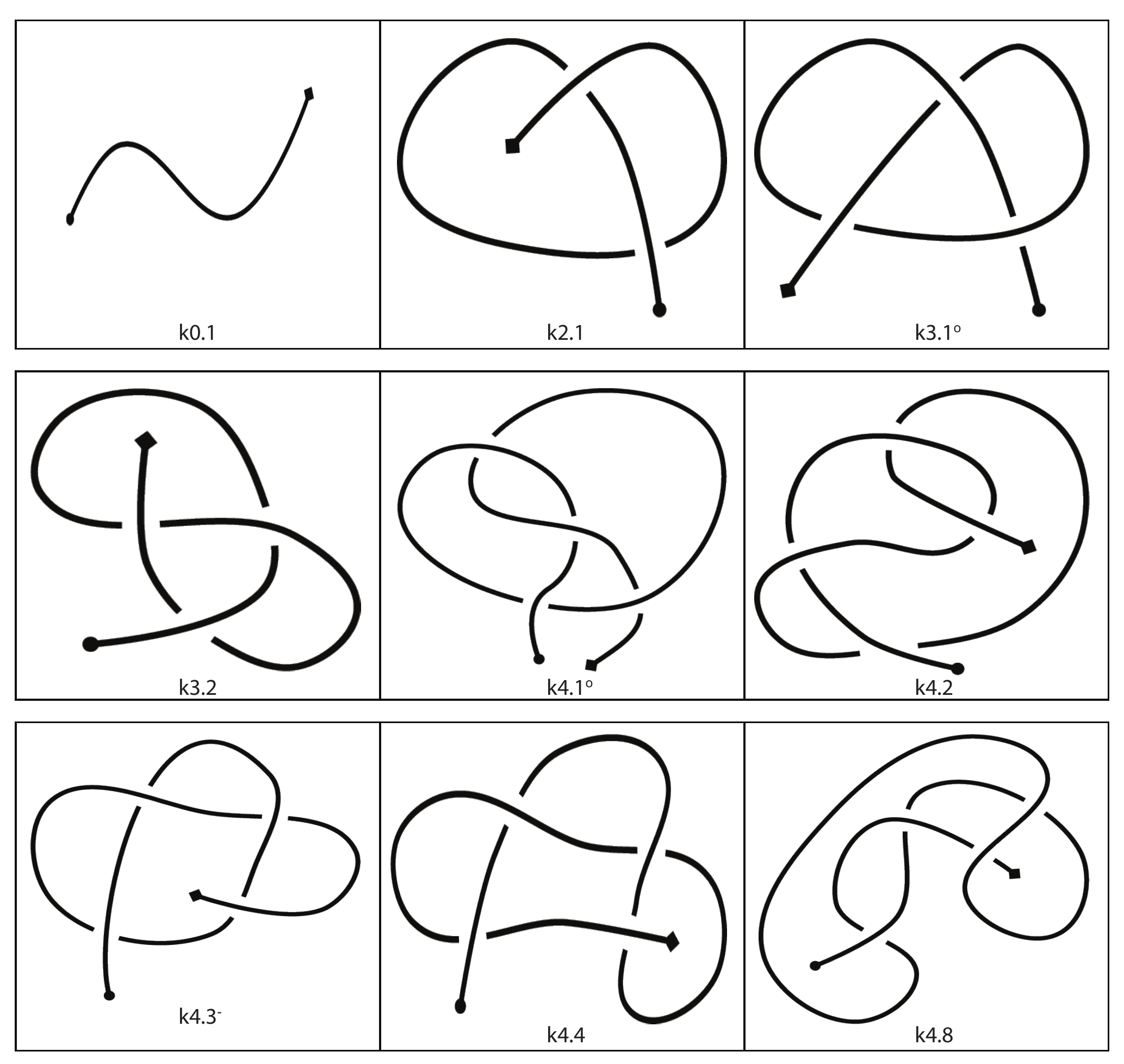}\
 \caption{Knotoid diagrams that appear in the present study.}
\label{fig:diag_list1}
    \end{figure*}

\begin{figure*}[!thbp]
  \centering
\includegraphics[width=.9\textwidth]{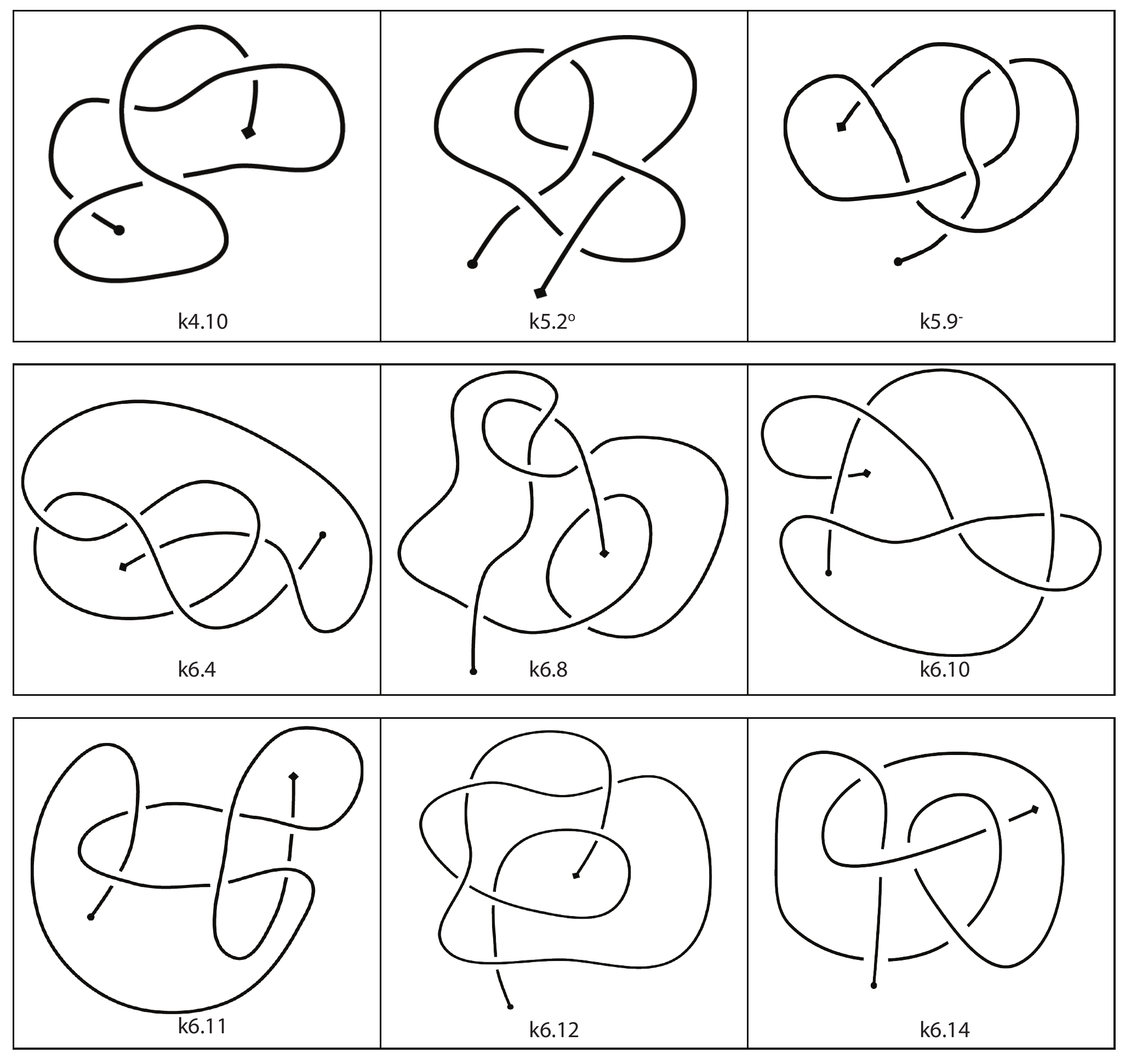}\
 \caption{Knotoid diagrams that appear in the present study.}
\label{fig:diag_list2}
    \end{figure*}
    
\pagebreak

\begin{figure}[!thbp]
  \centering
  \subfloat[1YVE]{\includegraphics[width=0.48\textwidth]{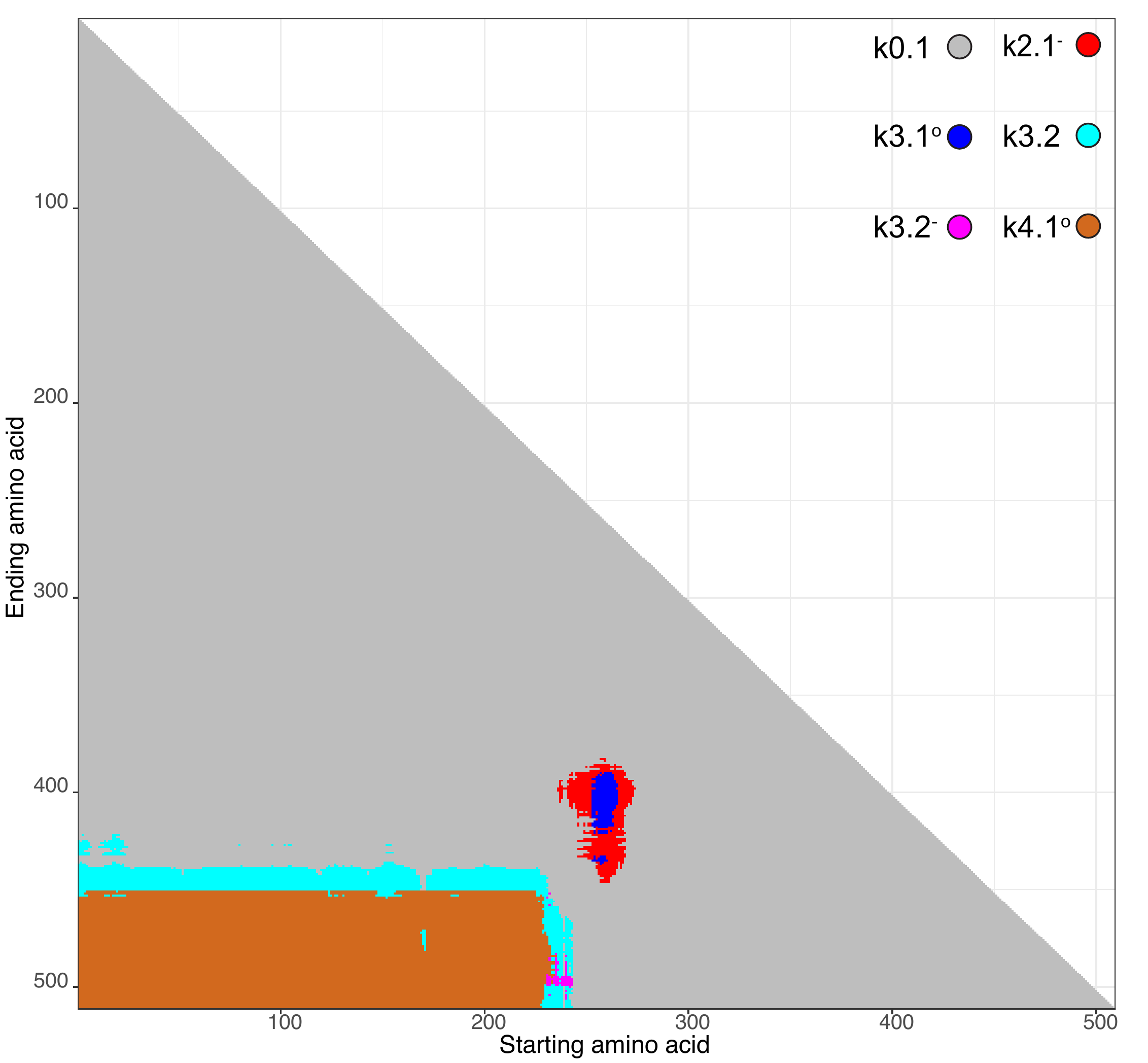}\label{fig:1YVE}}
  \hfill
  \subfloat[3FR8]{\includegraphics[width=0.48\textwidth]{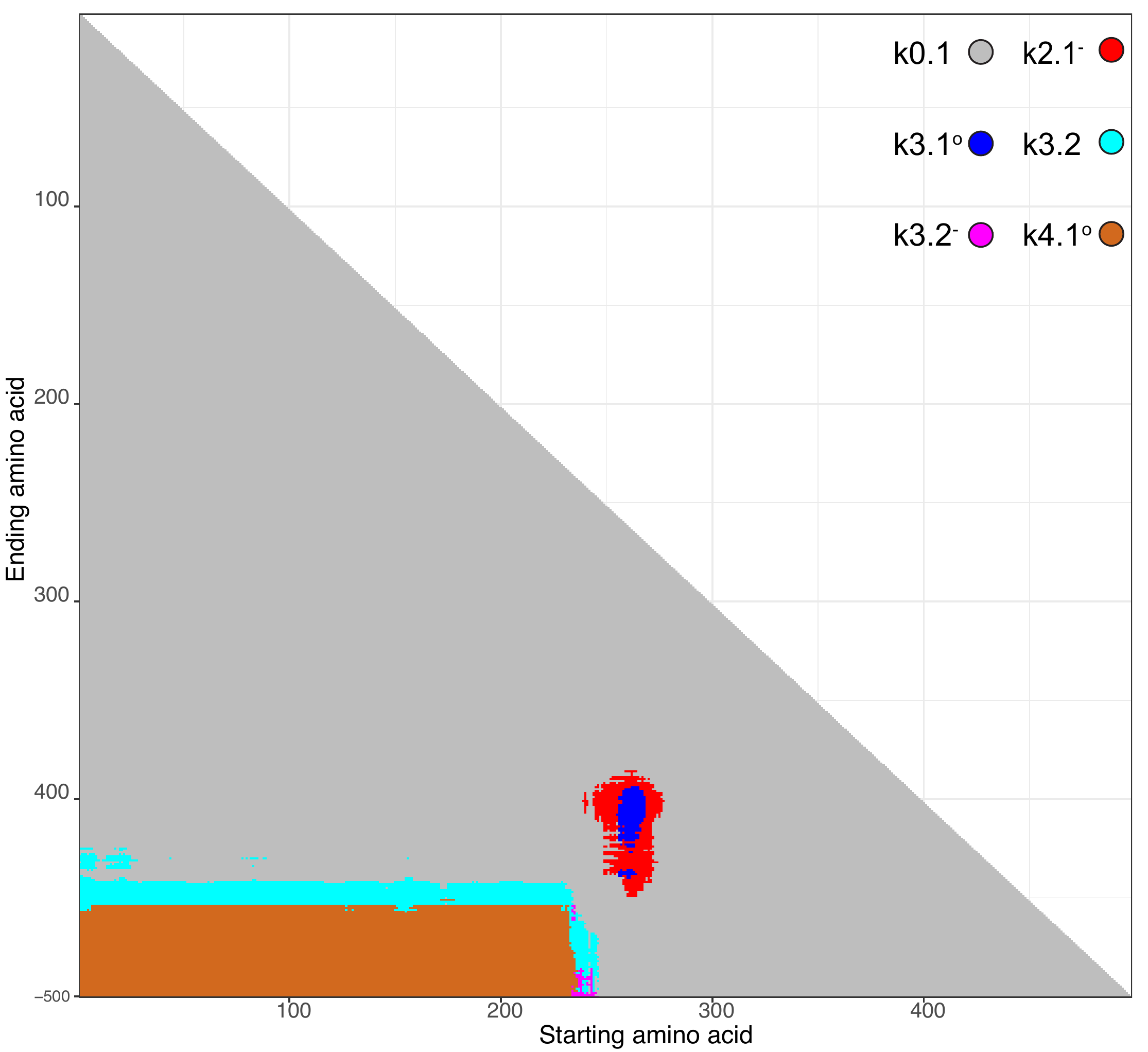}\label{fig:3FR8}} \\
  \subfloat[1YRL]{\includegraphics[width=0.48\textwidth]{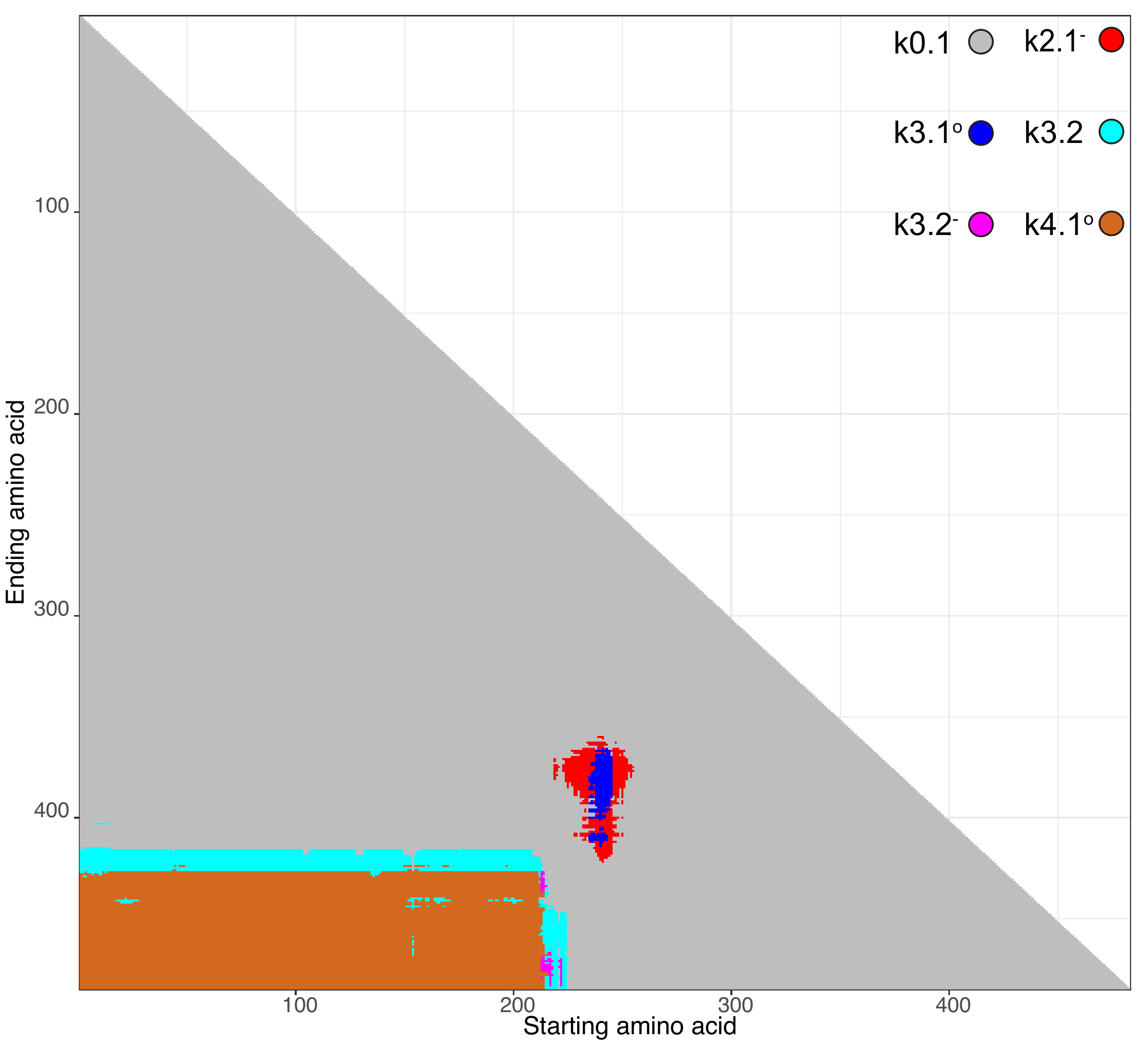}\label{fig:1YRL}}
  \hfill
    \subfloat[2AXC]{\includegraphics[width=0.45\textwidth]{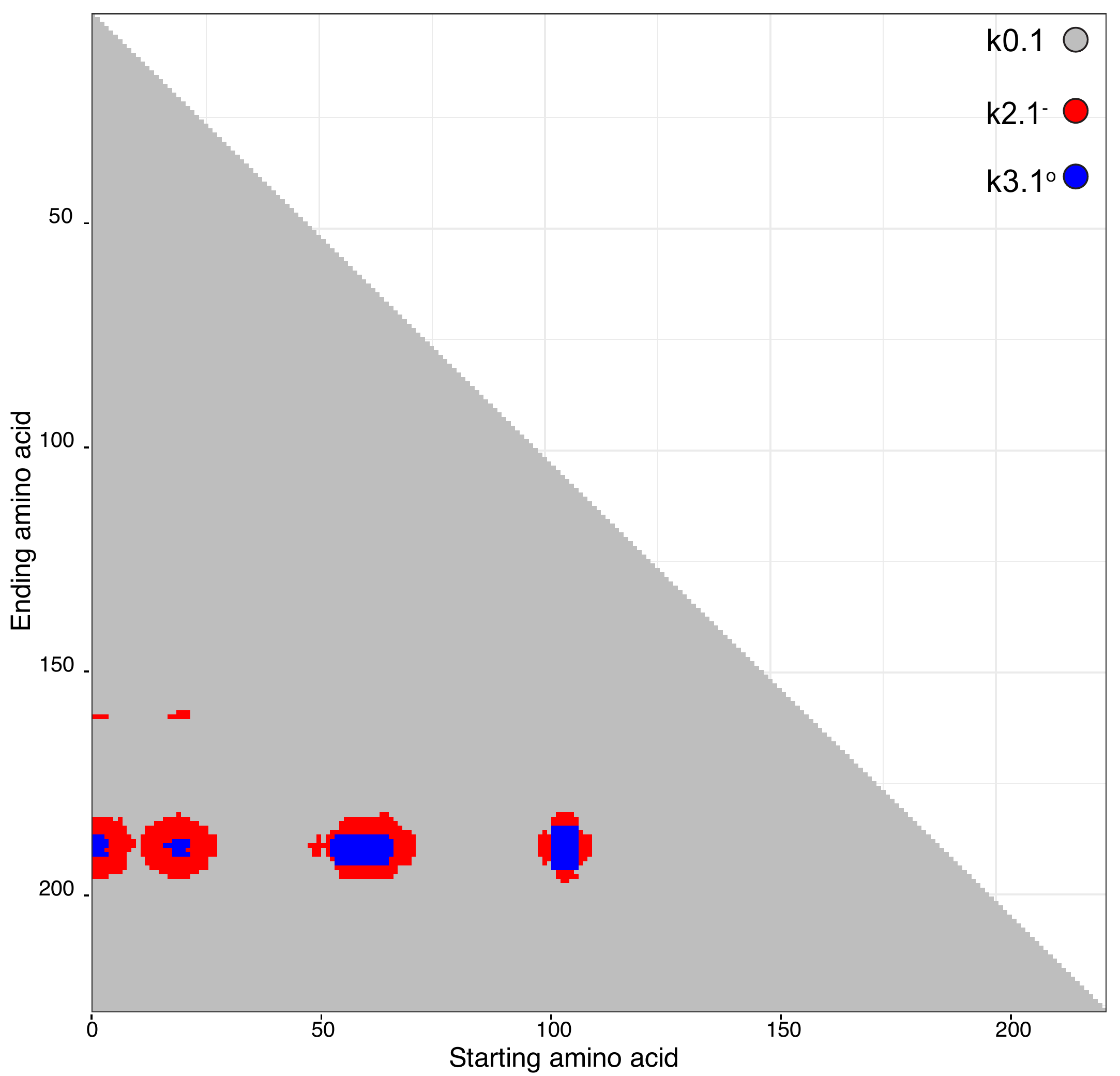}\label{fig:2AXC}}
   
      \caption{Knotoid fingerprints of protein chains 1YVE, 3FR8, 1YRL and 2AXC.}
    \end{figure}
    
 \begin{figure}[!thbp]   
\subfloat[3NCY]{\includegraphics[width=0.48\textwidth]{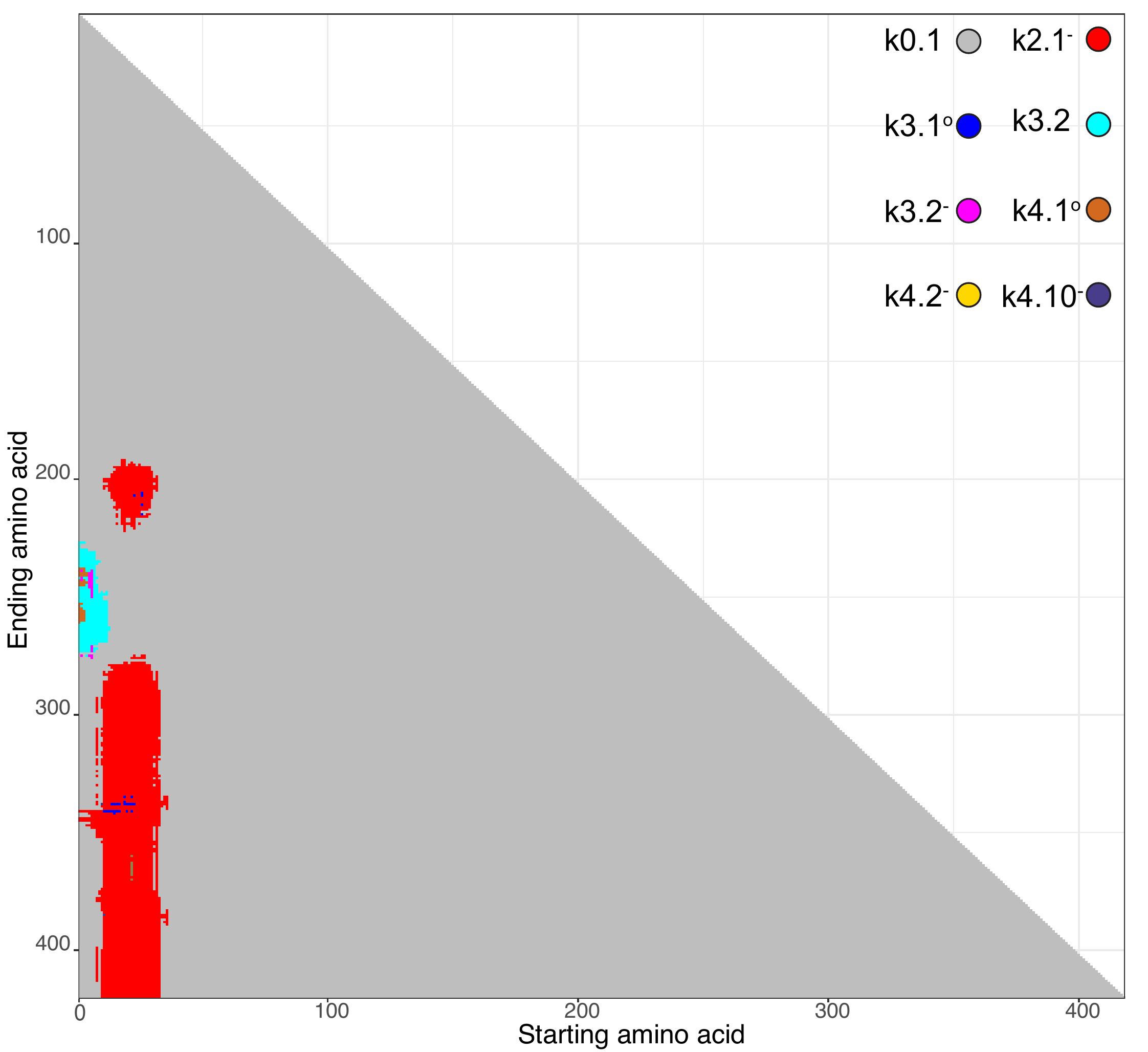}\label{fig:3NCY}}
\hfill
\subfloat[3L1L]{\includegraphics[width=0.48\textwidth]{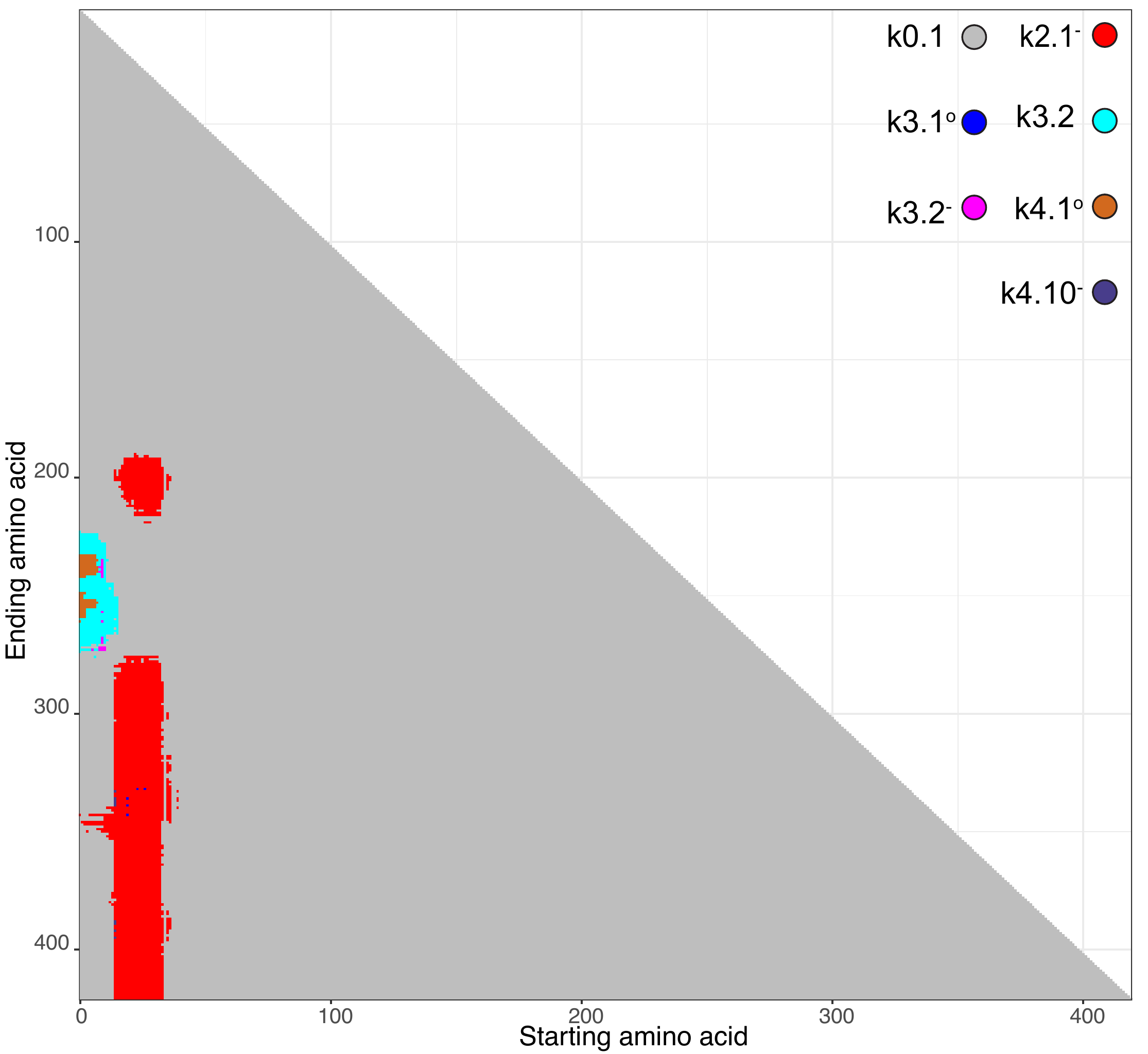}\label{fig:3L1L}} \\
 \subfloat[3DH4]{\includegraphics[width=0.48\textwidth]{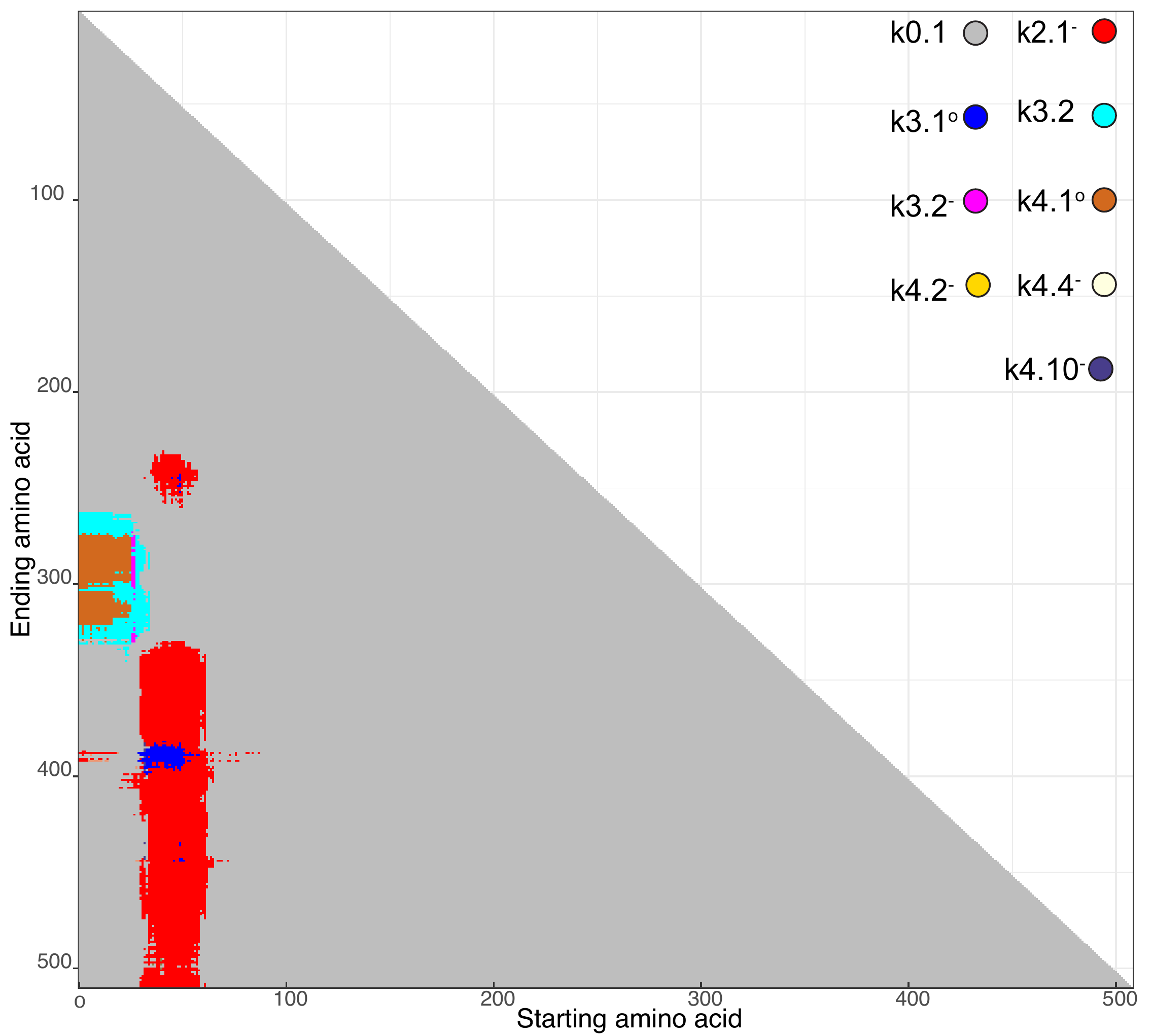}\label{fig:3DH4}}
   \hfill
  \subfloat[2JLO]{\includegraphics[width=0.48\textwidth]{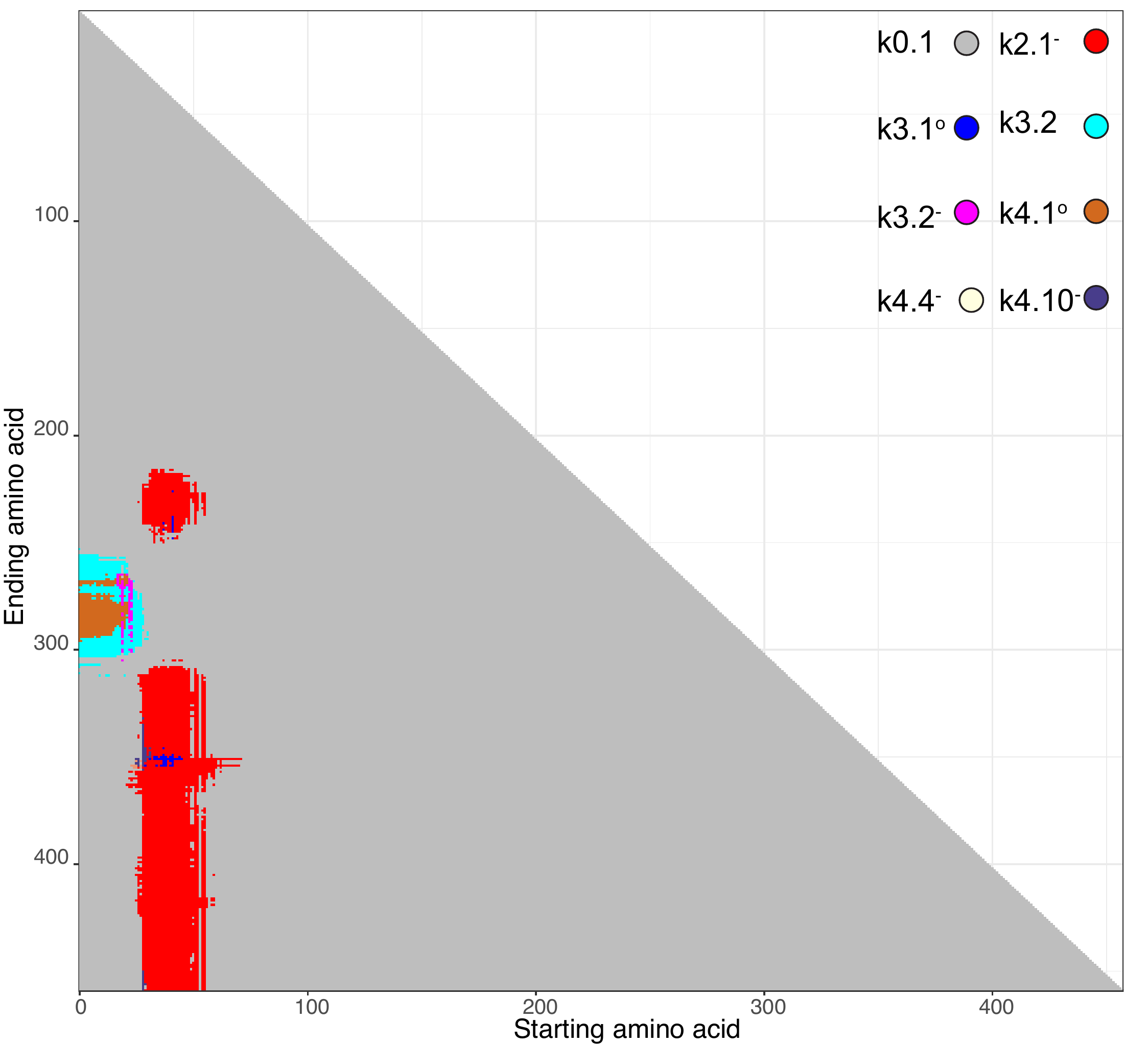}\label{fig:2JLO}}
  \caption{Knotoid fingerprints of protein chains 3NCY, 3L1L, 3DH4 and 2JLO.}
\end{figure}

 \begin{figure}[!thbp]   
  \subfloat[3IRT]{\includegraphics[width=0.48\textwidth]{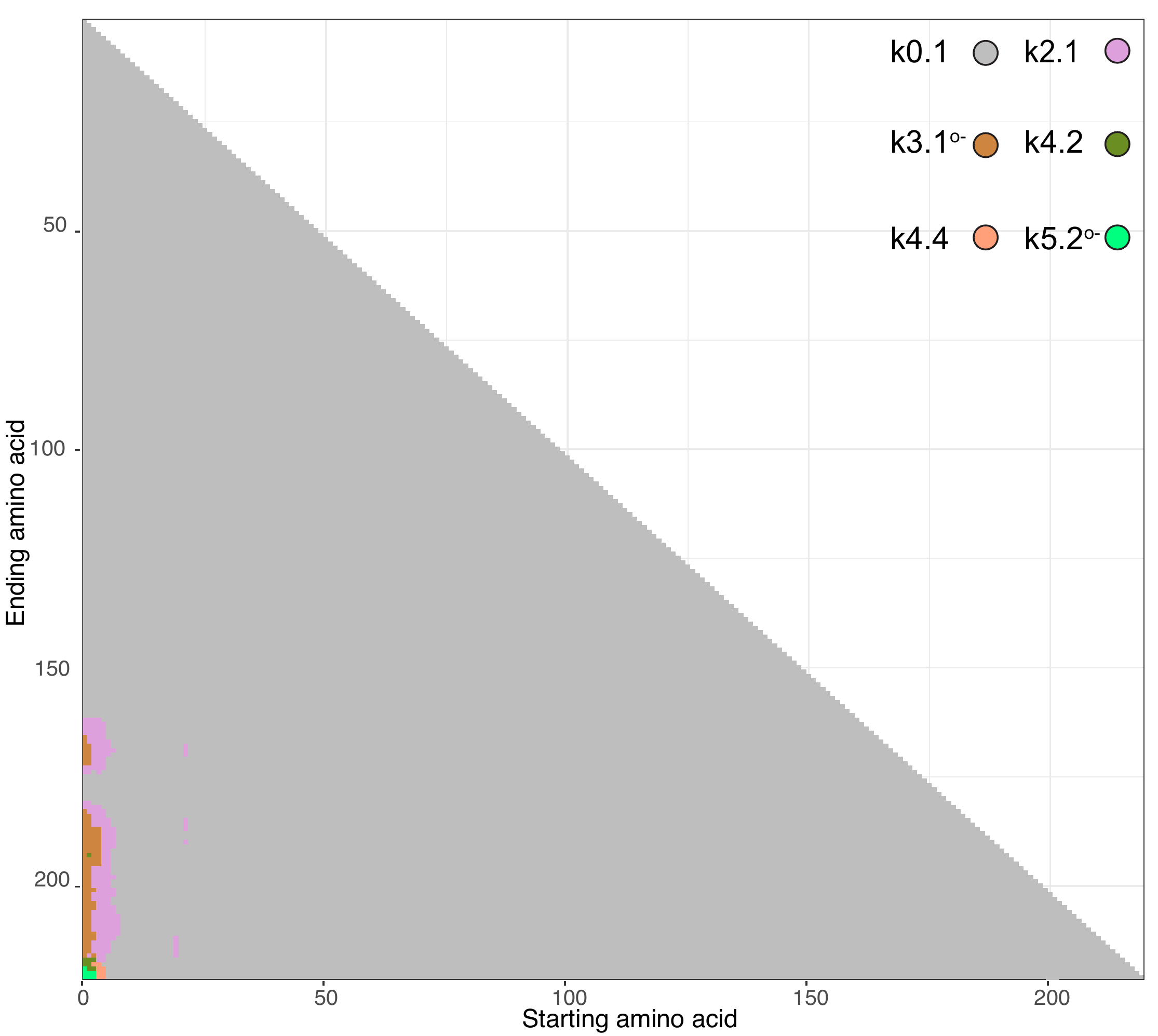}\label{fig:3IRT}}
  \hfill
   \subfloat[1XD3]{\includegraphics[width=0.48\textwidth]{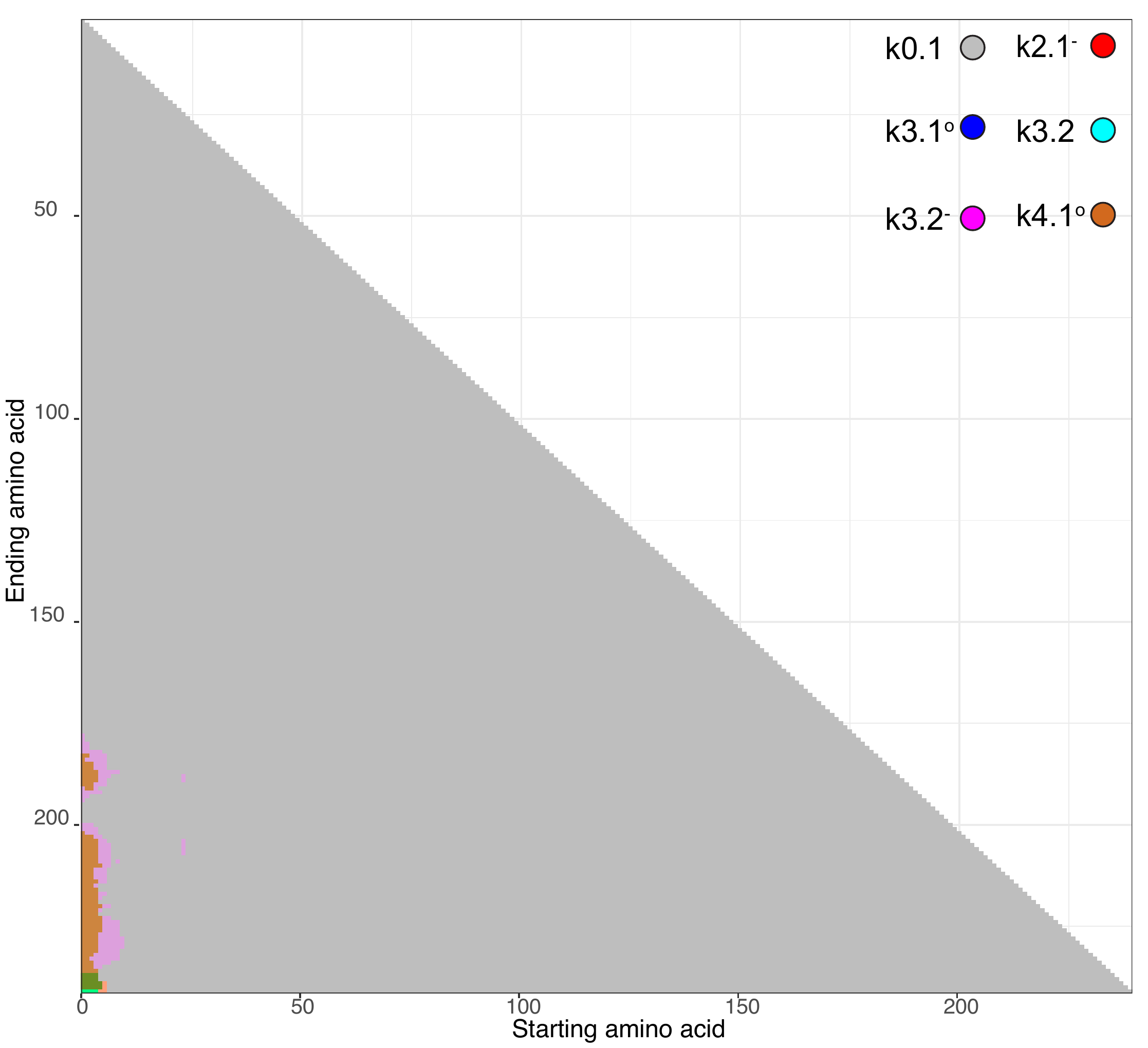}\label{fig:1XD3}}\\
  \subfloat[3C2W]{\includegraphics[width=0.48\textwidth]{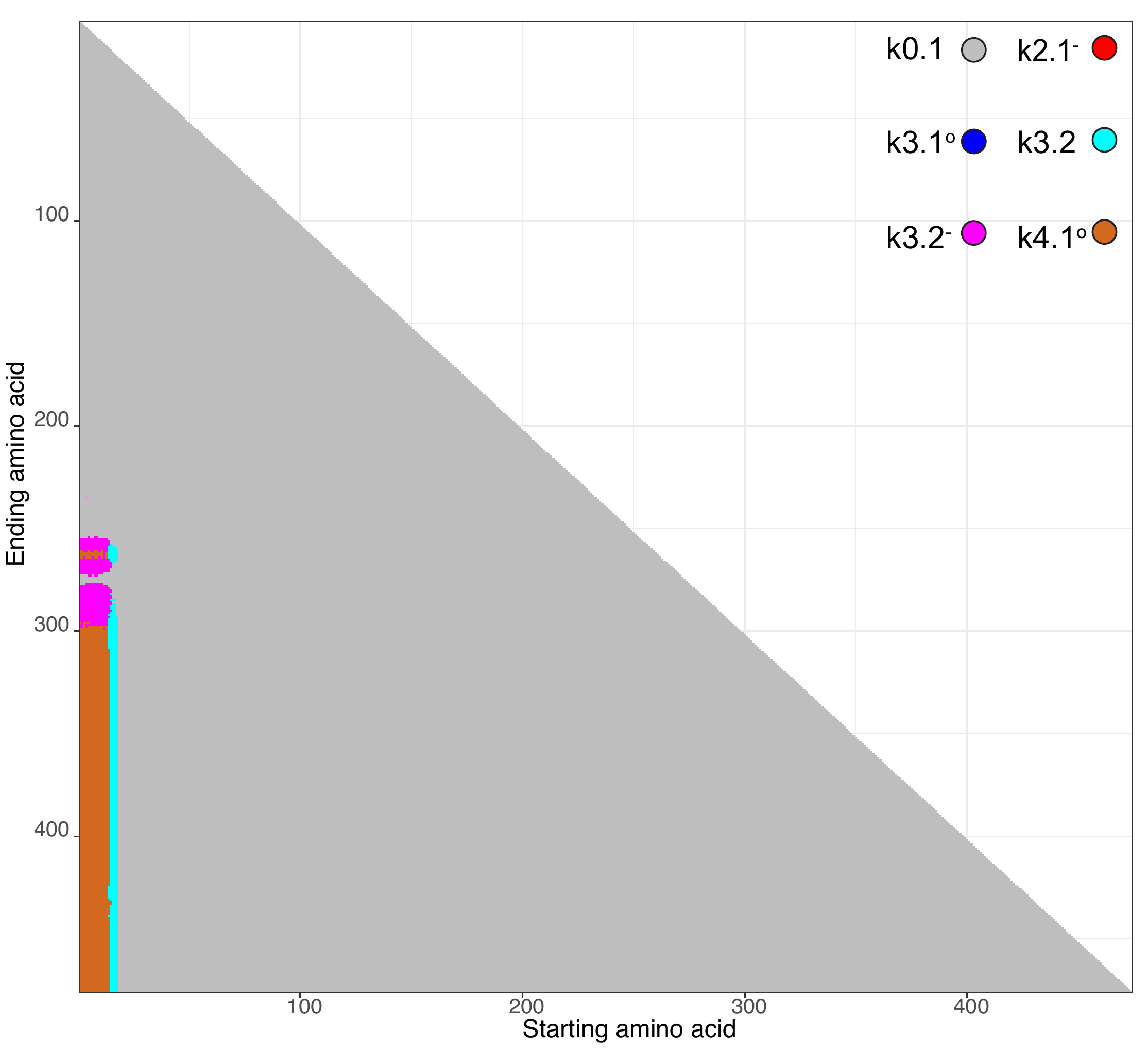}\label{fig:3C2W}} 
  \hfill
  \subfloat[2OOL]{\includegraphics[width=0.48\textwidth]{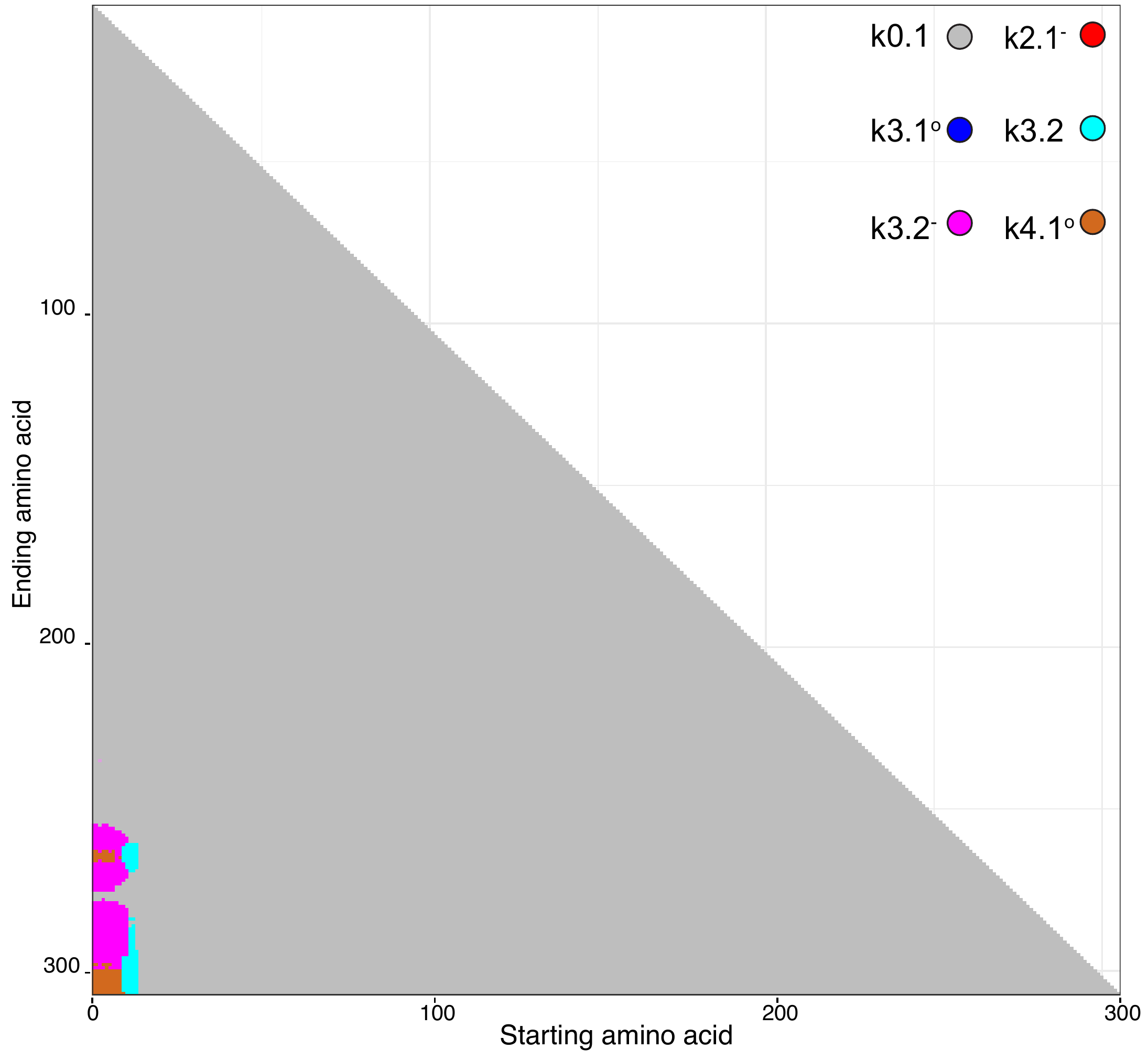}\label{fig:2OOL}}
   \caption{Knotoid fingerprints of protein chains 3IRT, 1XD3, 3C2W and 2OOL.}
\end{figure}

\end{document}